\documentclass{basi}
\usepackage{epsfig}
\begin{document}   

\title[]{Morphology and spectroscopy of hot gas in some early type galaxies}
\author[N.D.Vagshette, M.K.Patil, S.K.Pandey \& Ajit Kembhavi]{N.D.Vagshette,$^1$ M.K.Patil,$^1$ S.K.Pandey,$^2$ \& Ajit Kembhavi$^3$\\
	$^1$ School of Physical Sciences, SRTM University, Nanded-431 606, India\\
	$^2$ School of Studies in Physics, Pt. R. S. University, Raipur-492 010, India\\
	$^3$ Inter University Centre for Astronomy \& Astrophysics (IUCAA), Pune-411 007, India}
\maketitle

\begin{abstract}          
We present results of morphological and spectroscopic study of hot gas in some early-type galaxies based on the analysis of high resolution X-ray images acquired from the archive of Chandra space mission. Distribution of the hot gas in target galaxies after eliminating contribution from the discrete sources (LMXBs) displays varied morphologies, ranging from very compact nuclear emission to very extensive emission, larger than even optical images of the host galaxies. The surface brightness profile of the hot gas in program galaxies is well described by a single beta model, while spectrum of the diffuse emission is best fitted by a combined soft MEKAL model and a hard power law model. We use these results to derive temperature and abundance profiles of the hot gas in host galaxies. The deprojection of the diffuse emission shows a temperature gradient in some of the galaxies.

We also report on the 2-D distribution of the discrete sources (LMXBs) in host galaxies and compare it with their optical morphologies. The X-ray spectrum of the resolved sources is well-fit by a hard power law model with X-ray luminosities (0.3 to 10 keV) in the range from 5$\times$ 10$^{37}$ to 2.5$\times$ 10$^{39}$ erg s$^{-1}$. X-ray luminosity function (XLF) of the LMXBs shows a break near the luminosity comparable to the Eddington luminosity for a 1.4 M$_\odot$ neutron star.
\end{abstract}
\begin{keywords}
 Galaxies: elliptical and lenticulars, early-type; ISM:X-ray emission
\end{keywords}
\section{Introduction}
X-ray emission in early-type galaxies partly originate from the hot interstellar medium (ISM) which is distributed throughout the galaxy, and partly from the population of point-like sources, known as low mass X-ray binaries (LMXBs) (Trienchieri \& Fabbiano 1985). The spectra of the X-ray emission in these galaxies are typically fitted with a two-component model: a soft 1\,keV thermal component for hot, diffuse ISM and a hard 1.6 photon index power law component for the contribution from the point sources (Irwin et al. 2000). Though the diffuse X-ray emission is a powerful tool for tracing the past star formation and central massive black hole activity through their effects on the ISM (Strickland et al. 2000), the later component  is more important in the case of fainter galaxies (Matsushita 2001). The superb angular resolution of the {\em Chandra} X-ray telescope for the first time has made possible to spatially resolve the point sources in extragalactic environment. In this paper we present morphology and spectral properties of hot gas and point-like sources in some early-type galaxies.\\
\section{Observations and Data Analysis}
X-ray data on target galaxies NGC 3245, NGC 3607, NGC 3923 and NGC 4365 were acquired from  public archive of the {\em Chandra} X-ray observatory. Level-1 processed event files provided by the pipeline processing of {\em Chandra} X-ray centre were used for this study. The acquired X-ray data was homogeneously processed using CIAO (Chandra Interactive Analysis of Observations) science threads. A more detailed discussion on the X-ray data analysis is given in Strickland et al. (2002). 

The point sources present in the field of view centred on the target galaxies were identified and adaptively removed using the task {\em wavdetect} available within CIAO. The ``holes" due to this removal were filled by extrapolating the background counts using {\em dmfilth}. Finally, the ``filled-in" smooth, diffuse X-ray images were produced in different energy bands i.e., soft (0.3-1 keV), medium (1-2 keV) and hard (2-10 keV) using the task {\em csmooth}. Figure 1(a) shows one of such image delineating the detected X-ray point sources as well as the point sources within NGC 4365.

\begin{figure}[!h]
\centering
\includegraphics[width=6.5cm]{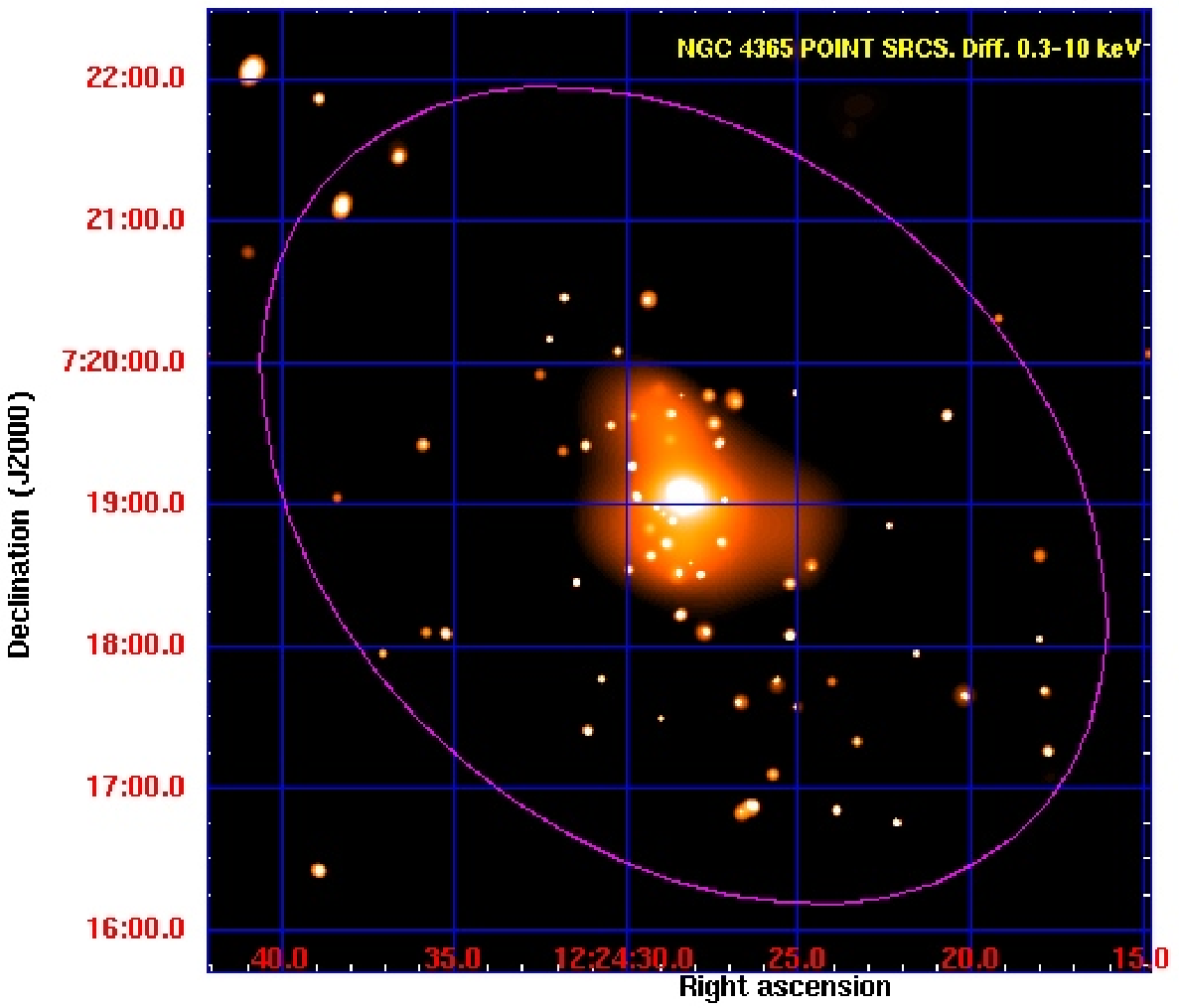}
\hspace{2mm}
\includegraphics[width=6.5cm]{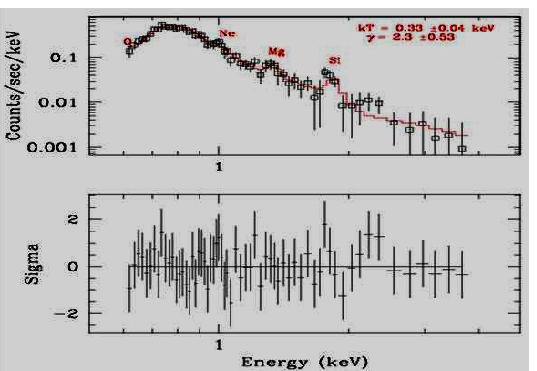} 
\caption{(a) Diffuse X-ray emission along with the detected point-sources in NGC 4365, elliptical contour represents $D_{25}$ region, (b) X-ray spectrum of hot gas alone in NGC 3923.}
\end{figure}

The spectra of the combined emission from the detected point sources with counts more than 100 net counts were fitted by a power-law spectrum with photon index about 1.6. These results are consistent with the spectral analysis of the combined emission from LMXBs in a sample of 15 early-type galaxies analyzed by Irwin et al. (2003). While, the spectra of the diffuse emission component in the target galaxies were fitted by a hot plasma code (VMEKAL) along with a power law function. Figure 1(b) represents X-ray spectrum of NGC 3923 for the diffuse gas alone. XSPEC V11.2 package was used for the spectral fitting.

\section{Results and Discussion}
We determined X-ray hardness ratios of the detected point sources following the technique discussed by Sarazin et al. (2000), which are useful for characterizing the spectral properties of the sources. Hardness ratios of all the detected sources in target galaxies are plotted as H31 vs. H21 and are shown in Figure 2(a), while Figure 2(b) represents the X-ray Luminosity Function (XLF) of the discrete point sources detected in the target galaxies.

 \begin{figure}[]
\centering
\includegraphics[width=5.5cm]{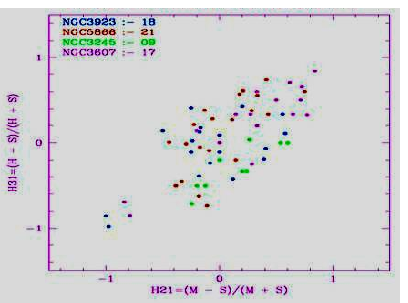}
\hspace{5mm}
\includegraphics[width=5.5cm]{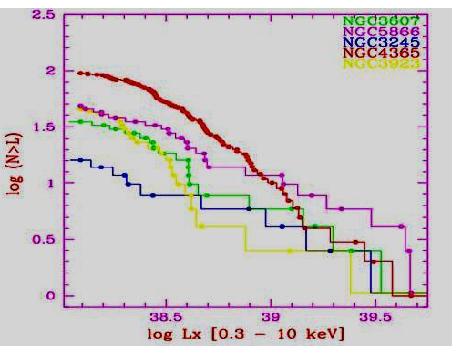}
\caption{(a)X-ray colour (Hardness ratio) plot of the point-sources detected in target galaxies, (b) X-ray Luminosity Function of the discrete point sources.}
\end{figure}

The spectrum derived for NGC 3923 (Figure 1(b)) shows lines from S, Si, Ne, Mg and Fe. We tried to determine the Si/Fe and S/Fe ratios by fitting a modified VMEKAL model to the diffuse gas component. All the other elemental abundances were allowed to vary freely, except He which was set to solar value. We find Z$_{Si}$/Z$_{Fe}$ ratio in NGC 3923 to be about 4.2, which is an indicator that the ISM is enriched by the supernova (SN) explosions.

The resulting best-fit power-law indices fitted for diffuse gas and detected point-sources are consistent with those derived from fitting the combined emission from the detected point sources. This shows that the spectral index of LMXBs does not vary significantly with the luminosity. Emission from LMXBs is significant in these galaxies and have a contribution lying between 30 to 65\% to the total X-ray emission of the target galaxies.

\hspace{2in} {\bf Acknowledgments}

We acknowledge the use of X-ray data from the public archive of {\em Chandra} observatory. This work is supported by ISRO, Bangalore under the RESPOND scheme, sanction No. ISRO/RES/2/334/2006-07.

\end{document}